\documentclass[twocolumn]{aastex631}

\usepackage{upgreek}
\usepackage[T1]{fontenc}

\shorttitle{NIRCam yells at cloud}

\begin{document}

\title{NIRCam yells at cloud: JWST MIRI imaging can directly detect exoplanets of the same temperature, mass, age, and orbital separation as Saturn and Jupiter}

\correspondingauthor{Rachel Bowens-Rubin}
\email{rbowens-rubin@mit.edu}

\author[0000-0001-5831-9530]
{Rachel Bowens-Rubin}
\affiliation{Department of Astronomy, University of Michigan, Ann Arbor, MI 48109, USA}
\affiliation{Eureka Scientific Inc., 2542 Delmar Ave., Suite 100, Oakland, CA 94602, USA}

\author[0000-0001-5864-9599]{James Mang}
\altaffiliation{NSF Graduate Research Fellow.}
\affiliation{Department of Astronomy, University of Texas at Austin, Austin, TX 78712, USA}

\author[0000-0002-9521-9798]{Mary Anne Limbach}
\affiliation{Department of Astronomy, University of Michigan, Ann Arbor, MI 48109, USA}

\author[0000-0001-5365-4815]{Aarynn L. Carter}
\affiliation{Space Telescope Science Institute, 3700 San Martin Dr, Baltimore, MD 21218, USA}

\author[0000-0002-7352-7941]{Kevin B. Stevenson}
\affiliation{Johns Hopkins APL, 11100 Johns Hopkins Rd, Laurel, MD 20723, USA}

\author[0000-0002-4309-6343]{Kevin Wagner}
\affiliation{Department of Astronomy and Steward Observatory, Univetsity of Arizona, 933 N Cherry Ave, Tucson, AZ 85721}

\author[0000-0002-1652-420X]{Giovanni
Strampelli}
\affiliation{Space Telescope Science Institute, 3700 San Martin Dr, Baltimore, MD 21218, USA}

\author[0000-0002-4404-0456]{Caroline V. Morley}
\affiliation{Department of Astronomy, University of Texas at Austin, Austin, TX 78712, USA}

\author[0000-0001-6831-7547]{Grant Kennedy}
\affiliation{Department of Physics, University of Warwick, Gibbet Hill Road, Coventry CV4 7AL, UK}

\author[0000-0003-0593-1560]{Elisabeth Matthews}
\affiliation{Max-Planck-Institut für Astronomie, Königstuhl 17, D-69117 Heidelberg, Germany}

\author[0000-0001-7246-5438]{Andrew Vanderburg}
\affiliation{Department of Physics and Kavli Institute for Astrophysics and Space Research, Massachusetts Institute of Technology, Cambridge, MA 02139, USA}

\author[0000-0002-5082-6332]{Ma\"issa Salama}
\affiliation{Astronomy Department, University of California Santa Cruz,
1156 High St, Santa Cruz, CA 95064, USA}

\begin{abstract}

NIRCam and MIRI coronagraphy have successfully demonstrated the ability to directly image young sub-Jupiter mass and mature gas-giant exoplanets. However, these modes struggle to reach the sensitivities needed to find the population of cold giant planets that are similar to our own Solar System's giant planets ($T_{\rm eff} = 60 - 125$\,K; $a=5 - 30$\,AU). For the first time, we explore the high-contrast imaging capabilities of MIRI imaging rather than coronagraphy. Using data from the JWST GO 6122: \textit{Cool Kids on the Block} program which targets nearby ($<6$\,pc) M-dwarfs with NIRCam coronagraphy and MIRI imaging, we demonstrate that 21\,$\mu$m MIRI imaging can detect planets with the same temperature, mass, age, and orbital separations as Saturn and Jupiter. For systems within 3\,pc, 21\,$\mu$m MIRI imaging reaches the sensitivity needed to detect planets colder than Saturn ($<95$\,K). NIRCam coronagraphy can achieve similar results only in the unlikely case that a cold giant planet is cloud-free. Motivated by these compelling findings, we extend our analysis to evaluate the measured performance of MIRI F2100W imaging versus NIRCam F444W coronagraphy to 70\,pc and conclude that MIRI imaging offers the advantage for systems within 20\,pc. Microlensing surveys predict an occurrence rate as high as 1 - 2 low-mass giant exoplanets per star, suggesting that JWST MIRI imaging surveys of nearby systems may be poised to uncover a substantial population. This breakthrough enables a path towards the first direct characterization of cold giant exoplanets that are analogous to the solar system giant planets.

\end{abstract}


\vspace{10mm}
\section{Introduction} \label{sec:intro}

Cold giant planets with temperatures, separations, and masses similar to those of the Solar System’s giant planets remain underrepresented among the known exoplanets. However, microlensing and radial velocity surveys have hinted that the population of cold giant planets may be common. For example, microlensing results from \citealt{Poleski2021} indicate that low-mass giant planets (star-planet mass ratios between $10^{-4} - 0.033$; semi-major axis of $5-15$\,AU) may be as common $1.4^{+0.9}_{-0.6}$ planets per system.  Long-baseline radial velocity surveys have indicated that giant planets less massive than Jupiter (0.1 - 1$M_{jup}$) are 
more prevalent than the population of larger gas giants ($>1 M_{jup}$) and that their orbital distribution may be clustered near or just beyond the  water snow line \citep{Fernandes2019,Fulton2021,Lagrange2023}. 
If these trends hold, low-mass cold giant exoplanets may be among the most common types of exoplanets in the galaxy.

The region near the system's water snow line is expected to play a key role in planet formation. Beyond this boundary, temperatures are low enough for compounds to condense into solid ice grains. This increases the building materials available and accelerates the process of building large planets.  
Most giant planets are believed to form outside the location of their primordial water snow line. 
Our four Solar System giant planets - Jupiter, Saturn, Uranus, and Neptune  - are consistent with this theory and are currently found in orbits outside the location of our own primordial snow line (2.7\,AU;  \citealt{Martin2012}) between 5.2 - 30\,AU. 
It is thus critical that we are able characterize the giant exoplanets on similar orbits as our own giant planets if we wish to place our Solar System in context with the broader diversity of planetary systems.

The direct imaging method offers the potential to discover and characterize these worlds. While microlensing can detect this population, the events are non-repeatable and thus preclude follow-up characterization. Radial velocity requires decades of monitoring to probe such wide orbits. Direct imaging is capable of revealing an exoplanet candidate and measuring its thermal emission and basic atmospheric properties in a single time epoch. With repeated observation, the direct imaging method can be used to validate planet candidates, measure the precise orbit of the planet, and study its variability.  
However, due to the challenges associated with ground-based direct imaging \citep{Chauvin2024, Follette2023}, the method has so far been limited to characterizing only the most massive ($>3 M_{jup}$), hottest ($>500$\,K), and youngest ($<500$\,Myr) gas-giant exoplanets.

JWST is proving transformative for expanding how we characterize previously known sub-stellar worlds and expanding the range of exoplanets accessible to be characterized with direct imaging and spectroscopy.
For example, disequilibrium chemistry was revealed in VHS 1256 b \citep{Miles2023}, CO$_2$ absorption and metallicity was found in HR 8799 c/d  \citep{Boccaletti2024, Balmer2025}, and enhanced metallicity and cloud signatures were seen in AF Lep b \citep{Franson2024}.
MIRI coronagraphy enabled the first direct detection of a mature giant planet in the Eps Indi system \citep{Matthews2024} as well as the   young low-mass giant planet embedded in the disk of TWA 7 \citep{Lagrange2025}. Four Cycle 3 programs currently aim to expand the census of sub-Jupiter mass exoplanets on wide orbits: GO 5825 \citep{Carter2024jwstprop}, GO 6012 \citep{Millar-Blanchaer2024jwstprop}, GO 6122 \citep{CoolKidsProp}, and SURVEY 6005.\footnote{6005 Survey: \url{https://www.stsci.edu/jwst/science-execution/program-information?id=6005}}

The majority of JWST direct imaging surveys searching for exoplanets observe using NIRCam coronagraphy with dual-band imaging in the F200W ($1.988 \pm 0.461\mu$m) and F444W ($4.401 \pm 1.023\mu$m) filters. However, models predict that cold ($<350$\,K) exoplanets exhibit clouds that can suppress flux in these NIRCam bands \citep{Burrows2003, Morley2014, Sudarsky2003}. High-metallicity atmospheres may further dim flux from 3.5–5\,$\mu$m due to increased CH$_4$, CO, and CO$_2$ absorption. This behavior is observed in our Solar System giants \citep{2016PASP..128a8005N} as well as in 
Eps Indi Ab and TWA 7\,b that appear bright in the mid-infrared but faint or undetectable near 4$\mu$m \citep{Matthews2024, Lagrange2025}.  These empirical cases highlight the importance of mid-infrared ($>10\,\mu$m) for identifying cold exoplanets.

In this Letter, we present the first performance measurements from GO 6122, \textit{Cool Kids on the Block: The direct detection of cold ice giants and gas giants orbiting young low-mass neighbors} \citep{CoolKidsProp}. 
GO 6122 is a JWST Cycle 3 program targeting cold ($<200$\,K) low-mass giant planets orbiting six nearby (2.4–6\,pc) low-mass stars. This program is the first exoplanet direct imaging program to combine dual-band NIRCam coronagraphy (F444W+MASK335R, F200W+MASK335R) with MIRI F2100W imaging.
Section~\ref{sec:obs} describes the observing configuration of GO 6122; Section~\ref{sec:analysis} outlines our methods for measuring the contrast sensitivity. In Section~\ref{sec:results}, we present cloudy planet models alongside the measured spectra of Jupiter and Saturn to evaluate the exoplanet detectability around Wolf 359 and EV Lac. 
Section \ref{sec:discussion} applies the GO 6122 sensitivity results to assess the limits of NIRCam and MIRI when searching for companions in systems as far away as 70\,pc. Section \ref{sec:conclusion} highlights the key takeaways.
This work provides a framework for optimizing JWST observations for detecting cold giant exoplanets akin to those in our own Solar System.

\section{Observations} \label{sec:obs}

The observations presented were completed by JWST on December 9 \& 10 UT 2024 as part of the GO 6122 program.  The JWST data presented in this article were obtained from the Mikulski Archive for Space Telescopes (MAST) at the Space Telescope Science Institute. The data for the observations can be accessed via \dataset[doi: 10.17909/23r9-q596]{https://doi.org/10.17909/23r9-q596}.

The NIRCam coronagraphy and MIRI imaging observations were attempted as a grouped-uninterruptible sequence of three systems: Wolf 359, EV Lac, and AD Leo.  However, the NIRCam coronagraphy observations of AD Leo were not completed because of a guide star failure due to a nearby bright star. We therefore limit the discussion in this work to a comparison of the sensitivities of NIRCam F444W coronagraphy mode and MIRI F2100W imaging completed for Wolf 359 and EV Lac. 
 A summary is provided in Table \ref{tab:stellarprop} of the properties of the Wolf 359 and EV Lac systems.

\begin{deluxetable*}{c c c}
\tablecaption{Properties of Observed  Systems}
\label{tab:stellarprop}
\tablehead{\colhead{Property} & \colhead{Wolf 359} &\colhead{EV Lac}  }
\startdata
RA J2000          & 10 56 28.92 $^{(a)}$            & 22 46 49.73$^{(a)}$             \\ \hline
DEC J2000         & +07 00 53.00 $^{(a)}$           & 44 20 02.37$^{(a)}$             \\ \hline
Distance          & $2.4086 \pm 0.0004$\,pc $^{(a)}$ & $5.0515 \pm 0.0006$\,pc $^{(a)}$ \\ \hline
Parallax          & $415.18 \pm 0.07$\,mas $^{(a)}$  & $ 197.96 \pm 0.02$\,mas $^{(a)}$ \\ \hline
Spectral Type     & solar-metallicity M6 $^{(b)}$   & M4.0V $^{(c)}$                  \\ \hline
Age & 0.1 - 1.5\,Gyr $^{(d)}$ & 0.025 - 0.3\,Gyr $^{(d)}$
    \\ \hline
F444W app. mag    & 5.58 $^{(f)}$                   & 4.88 $^{(f)}$                   \\ \hline
F2100W app. mag   & 5.23 $^{(f)}$                   & 4.65 $^{(f)}$                   \\ \hline
Exoplanets & 1 unrefuted candidate; see (d) and (g) & No candidates 
\enddata
\tablenotetext{}{\\		
(a)	\citealt{GaiaDR3};		\\
(b)	\citealt{Kesseli2019};		\\
(c)	\citealt{Lepine2013};		\\
(d)	\citealt{bowens-rubin2023};		\\
(e)	\citealt{Shkolnik2009};		\\
(f)	Calculated with the Phoenix Sol model  \citep{Husser2013} accessible online on the Stellar DataBase (\textsc{sdb}) website: \url{drgmk.com/sdb}	\\ 
(g) \citealt{Tuomi2019} names an unrefuted Wolf 359 b candidate exoplanet consistent with a super-Neptune on an 8\,yr orbit ($P_{orb} = 2938 \pm 436$\,d,  $a = 1.845^{+0.289}_{-0.258}$\,AU) \\
}
\end{deluxetable*}

\subsection{NIRCam Observing Configuration}

NIRCam coronagraphy observations were completed using dual-band imaging with the F444W and F200W filters in conjunction with the MASK335R coronagraphic mask. No dither pattern was used.
We selected the MEDIUM8 readout to reduce exposure times as compared to deeper readouts to avoid risks from cosmic ray hits. We elected 8 groups and 4 integrations to reach a total exposure time of 3041\,s (50.7\,min) through 36 images. The SUB320A335R subarray provided a field of view of 
20'' × 20'' in F444W.\footnote{SUB320A335R subarray field of view is documented here: \url{https://jwst-docs.stsci.edu/jwst-near-infrared-camera/nircam-instrumentation/nircam-detector-overview/nircam-detector-subarrays}}

\subsection{MIRI Observing Configuration}
The MIRI F2100W imaging mode was chosen in order to take 
advantage of JWST's sensitivity 
at the wavelengths that cold (80-200\,K) planets are expected to be the brightest and clouds do not suppress emission. 
The MIRI observations were completed with a 4-point cycling dither pattern. To avoid saturation, we used the SUB256 (28.2 x 28.2 arcsec) array with a FASTR1 readout pattern. 

Wolf 359 used 11 groups/int and 250 integrations/exp with 1000 integrations for a total exposure of 3593\,s (59.9\,min). EV Lac used 7 groups/int and 375 integrations/exp with 1500 total integrations for a total exposure time of 3593\,s (59.9\,min). AD Leo used 5 groups/int and 400 integrations/exp with 1600 total integrations a total exposure time of 2874\,s (47.9\,min).

\section{Data Analysis} \label{sec:analysis}

\subsection{NIRCam Contrast Curve}

The NIRCam data were reduced using the \texttt{spaceKLIP} pipeline \citep{Kammerer2022, Carter2023_hip65426}. \texttt{spaceKLIP} serves as a wrapper to the default JWST pipeline\footnote{JWST pipeline code:\url{https://github.com/spacetelescope/jwst}} with additional steps for pseudo-reference pixel correction, 1/f noise subtraction, outlier detection, and image alignment. It also includes JWST-specific routines for PSF subtraction, contrast curve estimation, and companion astrometry and photometry. This analysis uses pipeline calibration version 1.16.1 and CRDS context \texttt{jwst\_1303.pmap}.

We started with the Stage 0 \texttt{*uncal.fits} files for Wolf 359 and EV Lac, processing them through \texttt{spaceKLIP} with the dark current correction skipped following \citet{Carter2023_hip65426}. A four-pixel boundary at the image edge is used as pseudo-reference pixels, jump detection thresholds are set to 4, and all pixels neighboring saturated pixels are flagged as saturated. 
We then apply a four-step pixel cleaning process: (1) DQ-flagged pixels are replaced via 2D interpolation (6-pixel window); (2) 11 poorly corrected DQ pixels are manually reflagged and interpolated again; (3) temporal 4$\sigma$ outliers are replaced with the pixel median; and (4) spatial 4$\sigma$/3$\sigma$ positive/negative outliers (3-pixel window) are replaced using the same 2D interpolation method.

To align the images for PSF subtraction, we first measured each target’s offset from the expected coronagraphic center via cross-correlation with a \texttt{WebbPSF} model \citep{Perrin2014}.  EV Lac and Wolf 359 yielded different offsets, so we reduced the data along two separate paths using the offset for each star. 
In each case, the non-centered target is co-aligned by applying the measured spatial offset. We then subtract the NIRCam background following \citealt{Lawson2024} and pad each image with an 80-pixel NaN border to prevent edge effects during PSF subtraction.

\begin{figure*}
    \centering
    \includegraphics[width=0.75\linewidth]{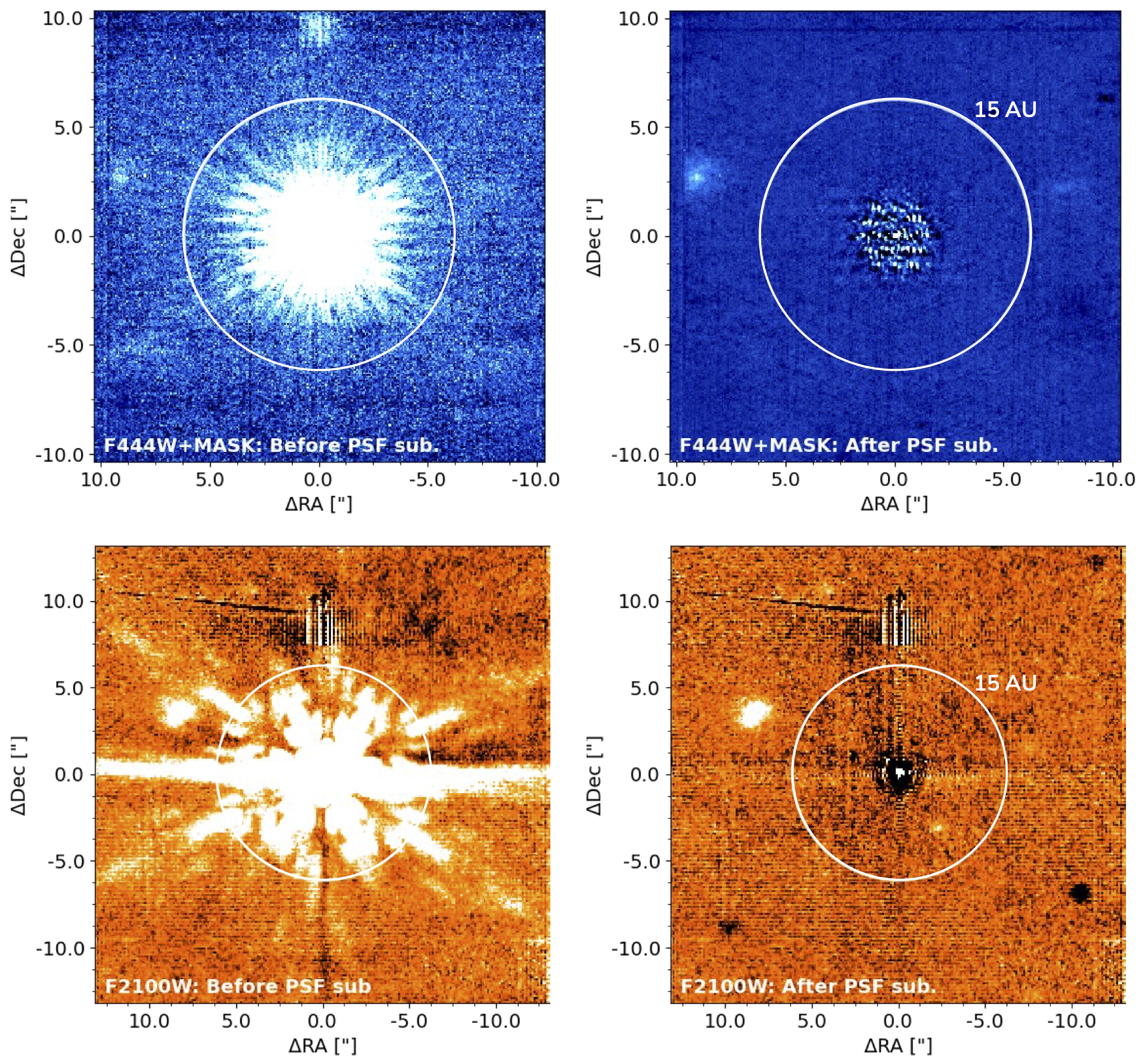}
    \caption{\textbf{Wolf 359 images from NIRCam (blue) and MIRI (red) before and after PSF subtraction.} PSF subtraction was performed in order to improve the sensitivity to finding faint astrophysical sources near the star. The white circle represents an angular separation of 15 AU for the Wolf 359 system.  The dark spots in the background of the PSF subtracted frames are due to positive sources in the reference star frames; this issue is expected to be mitigated in future analysis when additional reference stars are available. In the MIRI F2100W frames, an image artifact is present approximately 75 pixels above the center of the star.  Future proper motion followup observations are needed to identify if uncataloged point sources are co-moving companions. }
    \label{fig:NIRCamraw}
    \label{fig:MIRIraw}
\end{figure*}

PSF subtraction was carried out with \texttt{spaceKLIP} implementation of the  \texttt{pyKLIP} pipeline \citep{PyKLIP}, employing a single annulus and subsection with KL modes ranging from 1 to 40. Since there are only two targets, each star serves as the PSF reference for the other. No significant improvement is observed beyond the first KL mode as expected, given JWST’s thermal stability and precise pointing that minimize the variation in the PSF between integrations. The KL mode of 1 subtraction is shown for Wolf 359 in Figure \ref{fig:NIRCamraw}.

The contrast curves were calibrated following \citep{Carter2023_hip65426}. We included the effects of small-sample statistics, coronagraphic mask throughput, and KLIP throughput before  converting to physical units using the stellar fluxes in the JWST filters estimated from the best-fit spectral energy distribution (SED) models. The contrast curves are shown for the NIRCam F444W observations in the top row of Figure \ref{fig:cc}.  
We found that Wolf 359 and EV Lac align in the background-limited regime at a contrast of $6\times10^{-7}$ by 2.6 arcsec.  This contrast corresponds to an apparent magnitude sensitivity of $m_{F444} =  21.5$ for Wolf 359 and $m_{F444} =  20.8$ for EV Lac, as Wolf 359 is inherently fainter than EV Lac. At these wide separations, the flux of the star is negligible, and detector noise and background noise are dominant. As the exposure settings are identical between targets, the background noise must drive the difference in the apparent magnitude sensitivity between the objects. 

 Naively, it would be expected that EV Lac has a higher background level, as it has the worse magnitude sensitivity, however the reality is more counterintuitive as the measured background level for Wolf-359 is $\sim$20 times higher than that of EV Lac. The important aspect is the nature of our subtraction, where Wolf 359 and EV Lac are flux-scaled to be used as references for one another. As EV Lac is brighter, it is scaled down to match the flux of Wolf 359 prior to subtraction, and the residual noise from the background subtraction is also scaled down. Conversely, Wolf 359 is fainter and must be scaled up to match the flux of EV Lac, amplifying the residual noise by the flux scale factor and leading to its comparably worse magnitude sensitivity. Upon completion of the full 6122 program and through implementation of a broader PSF library, it may be possible to identify more suitable references for EV Lac and improve the sensitivity in the background regime. Alternatively, a more detailed exploration of subtraction strategies which optimize over distinct sub-divisions of the image may offer improvements.

\subsection{MIRI Contrast Curve}

The MIRI data were reduced using a custom pipeline with the data products accessible through the MAST Portal.\footnote{MAST: \url{https://mast.stsci.edu/portal/Mashup/Clients/Mast/Portal.html}} To address the non-uniform background structure with the MIRI flat-field, we performed the Stage 2 data reduction using a custom package, \texttt{MAGIC}, 
which is publicly available on GitHub.\footnote{\texttt{MAGIC}: \url{https://github.com/kevin218/Magic}} We used \texttt{Magic} Version 1.0 alongside JWST pipeline version 1.15.1 and CRDS jwst\_1225.pmap.
The background subtraction procedure followed an approach adapted from an STScI JWebbinar demo,\footnote{JWebbinar demo: \url{https://github.com/spacetelescope/jwebbinar_prep/blob/jwebbinar31/jwebbinar31/miri/Pipeline_demo_subtract_imager_background-platform.ipynb}} which applied a custom method utilizing multiple dithers per source.

To subtract the stellar flux, we applied a reference-star differential imaging strategy. For Wolf 359, we used the MIRI F2100W images of AD Leo and EV Lac as references; for EV Lac, the reference images were from AD Leo and Wolf 359. PSF subtraction was performed using the built-in principal component analysis (PCA) algorithms in the \emph{VIP: Vortex Imaging Processing} Python package (\texttt{VIP}) \citep{VIP}, version 1.6.4. Prior to subtraction, the frames were cropped to 240 pixels and NaN corrected. Image centering was completed using a positive Gaussian fit with the \texttt{ndimage-fourier} library. Figure~\ref{fig:MIRIraw} (bottom) shows the MIRI F2100W images before and after PSF subtraction.

\begin{figure*}
\gridline{\fig{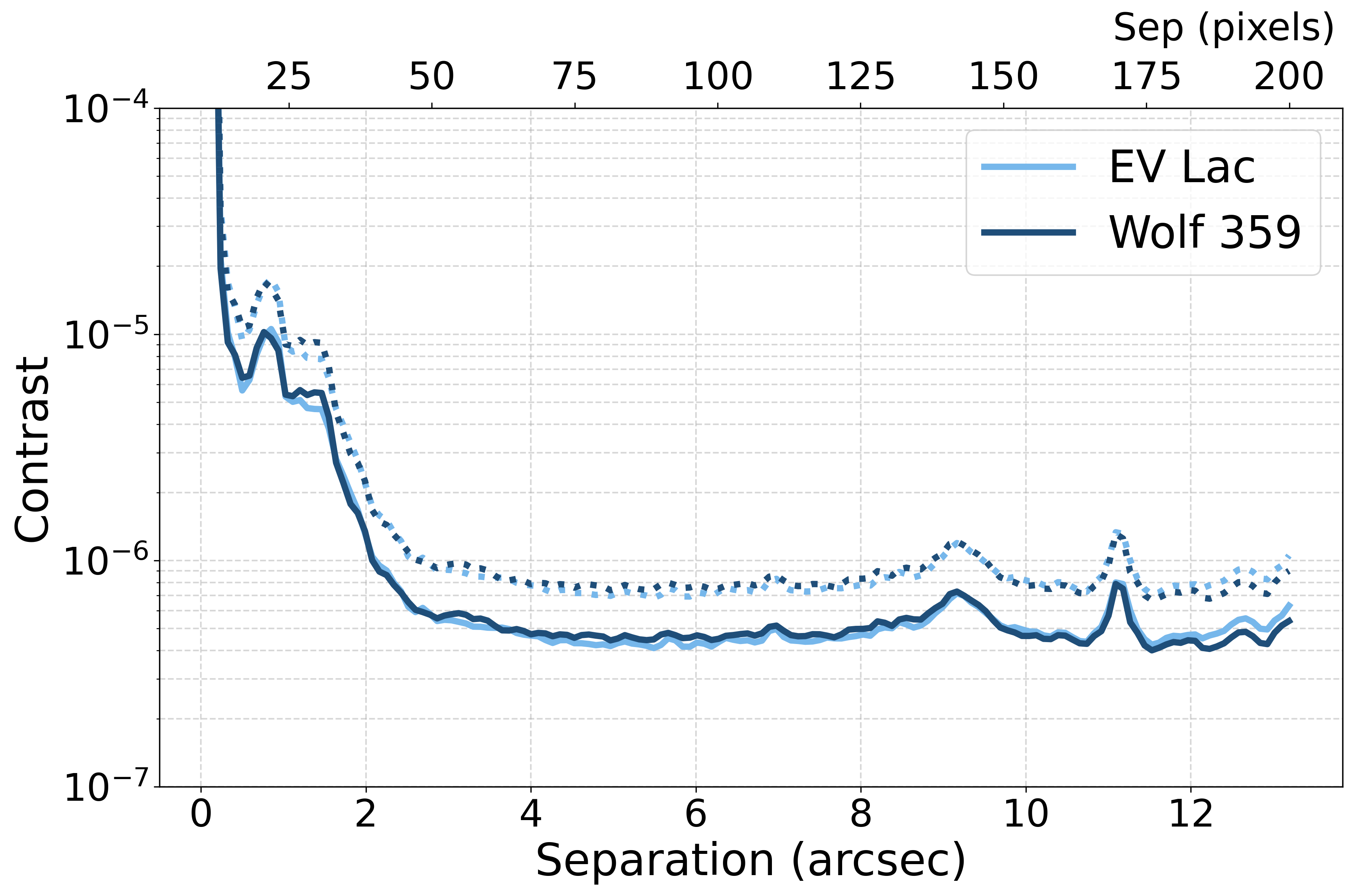}{0.4\textwidth}{(a) NIRCam F444W+MASK Coronagraphy Contrast}
          \fig{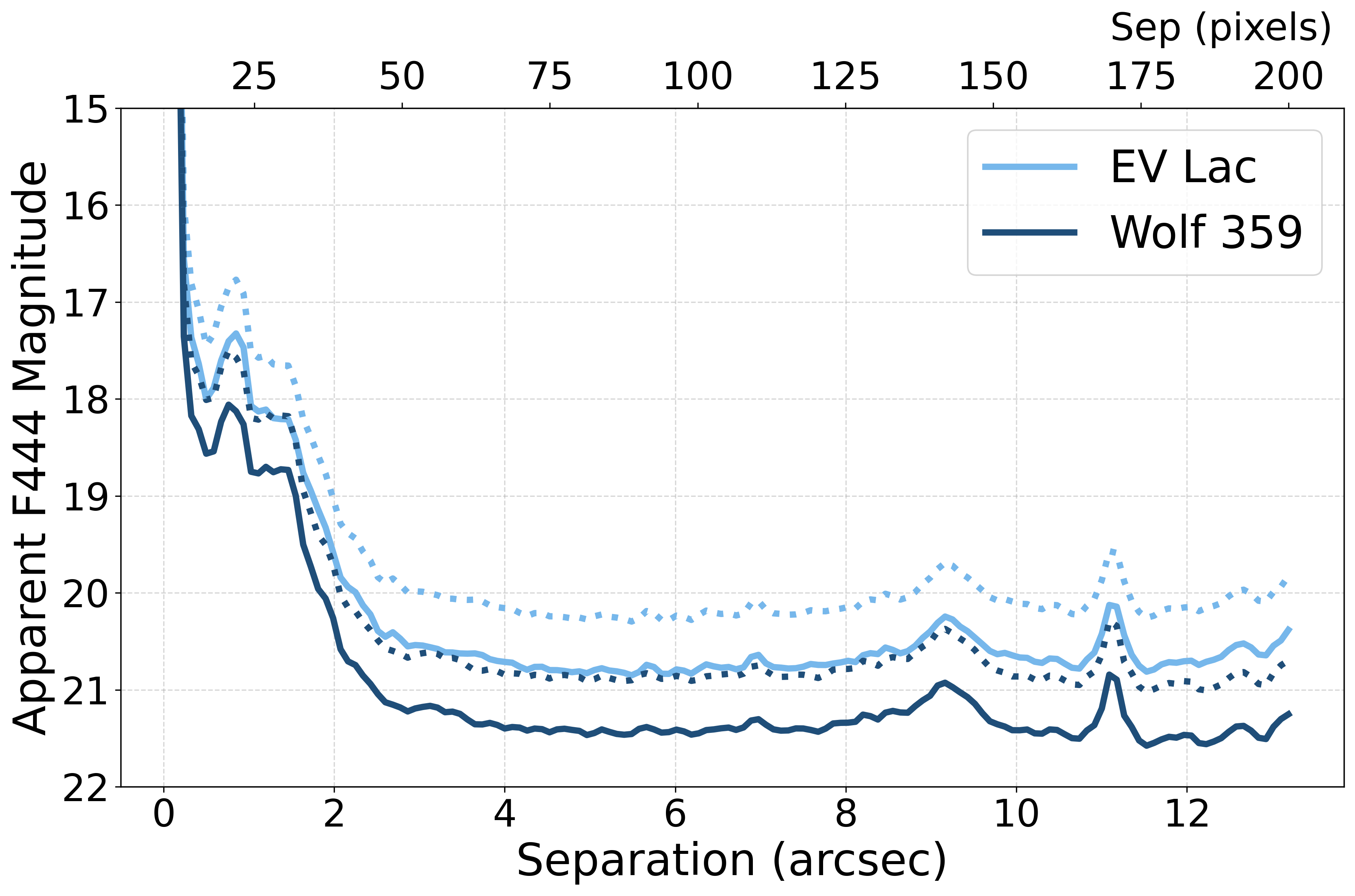}{0.4\textwidth}{(b) NIRCam F444W+MASK Coronagraphy Sensitivity in Apparent Magnitude}}
\gridline{\fig{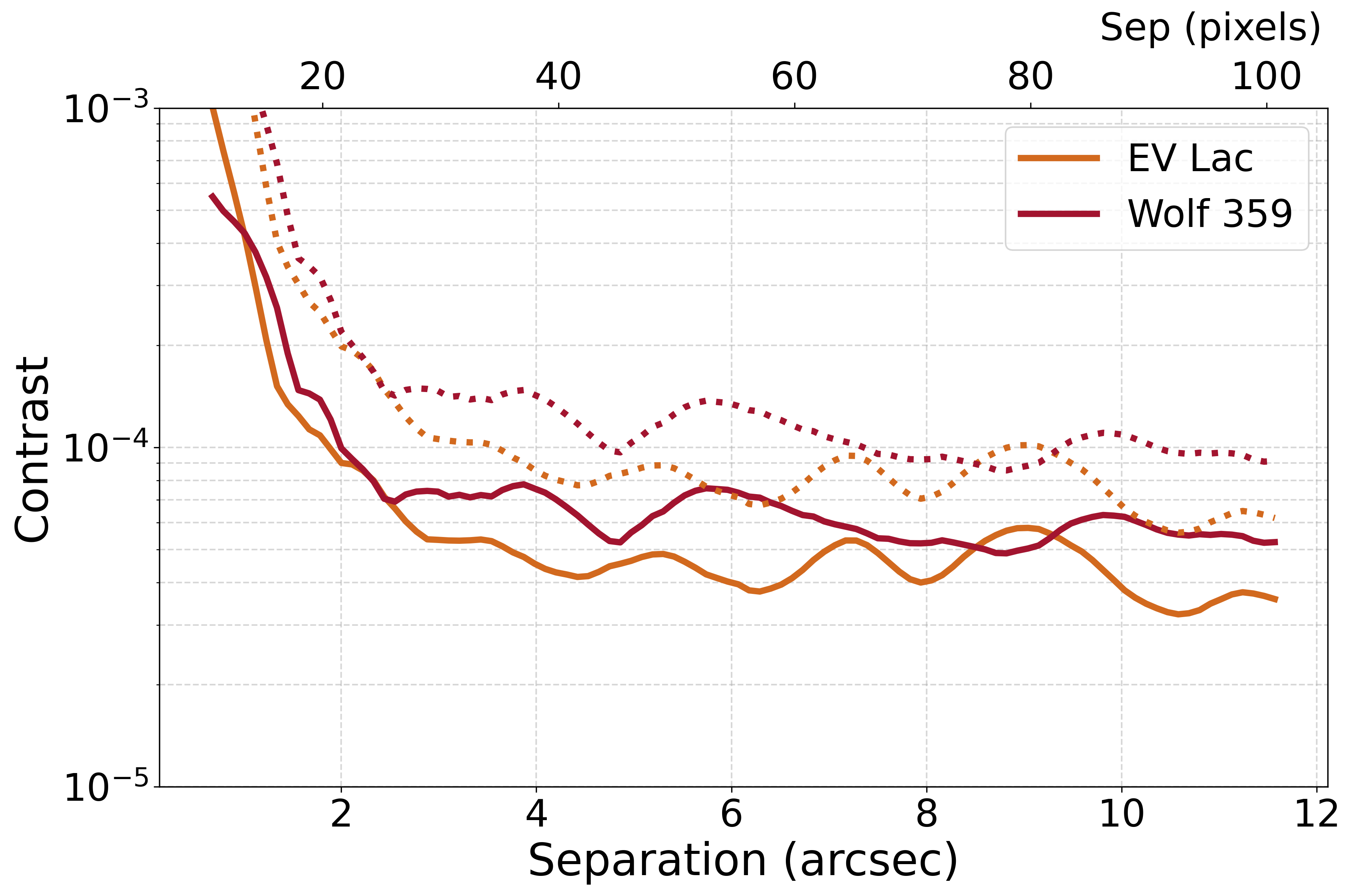}{0.4\textwidth}{(c) MIRI F2100W Imaging Contrast}
          \fig{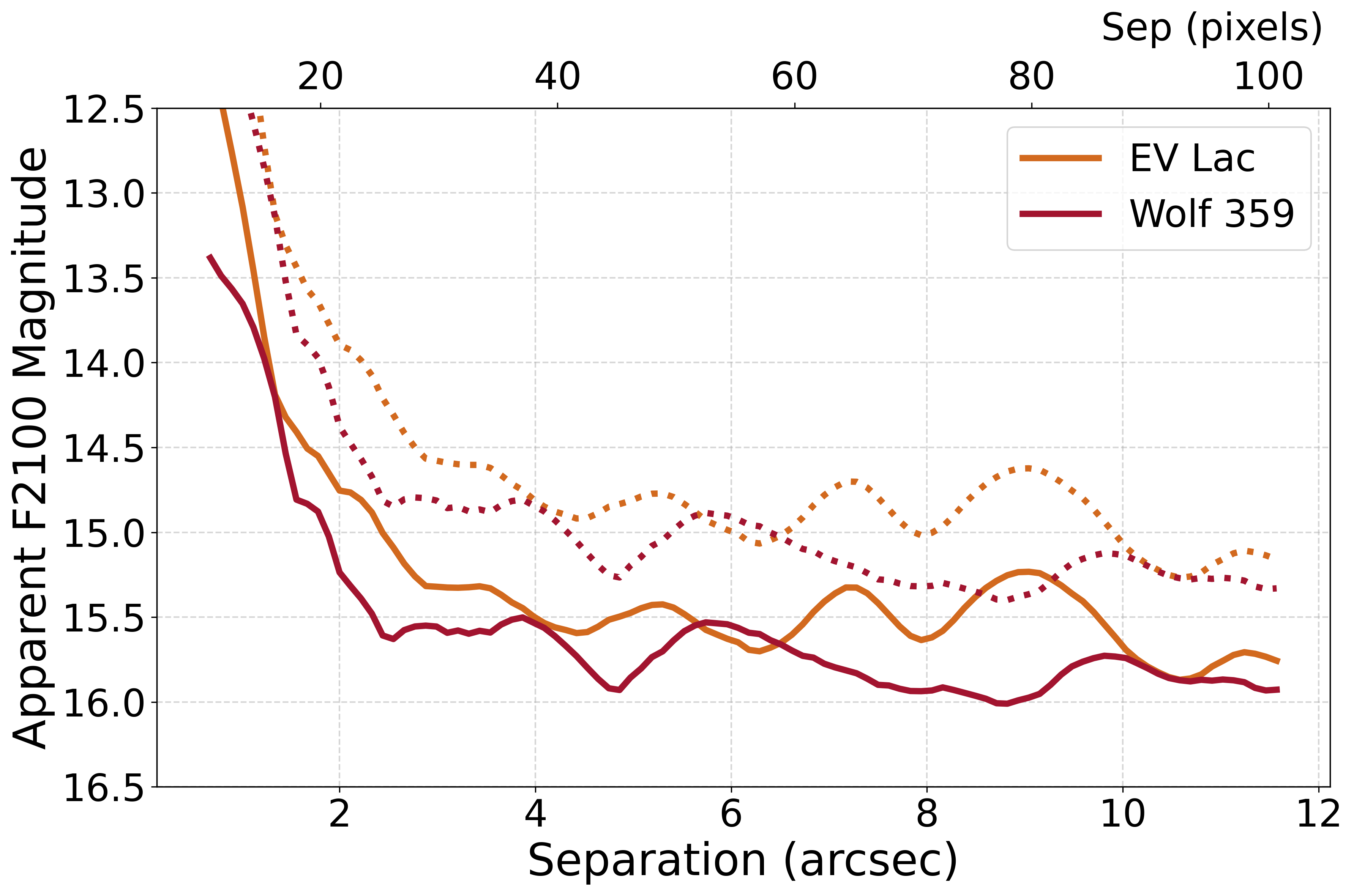}{0.4\textwidth}{(d) MIRI F2100W Imaging Sensitivity in Apparent Magnitude}}
\gridline{\fig{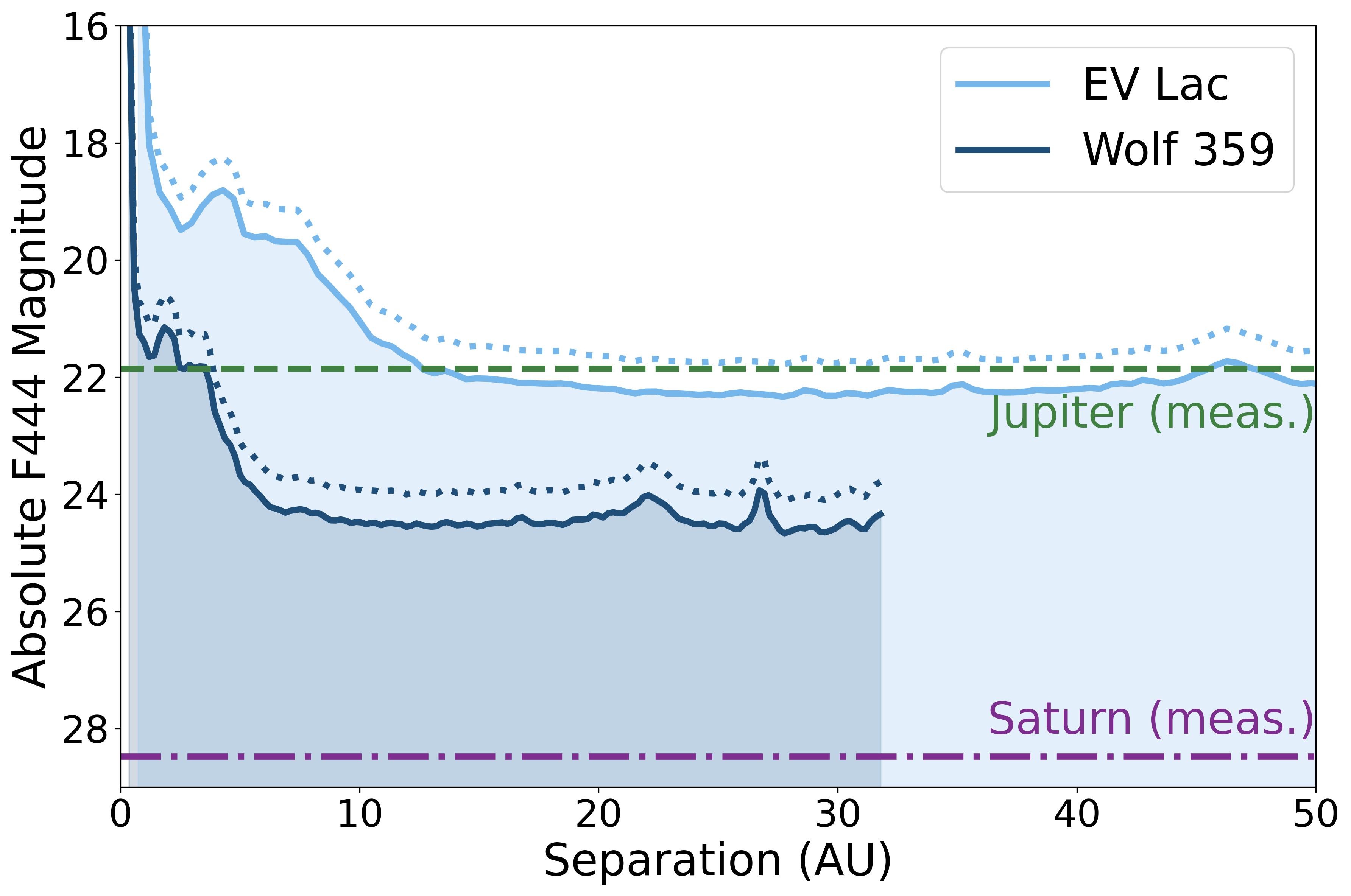}{0.4\textwidth}{(e) NIRCam F444W+MASK Coronagraphy Sensitivity in Absolute Magnitude versus AU}
          \fig{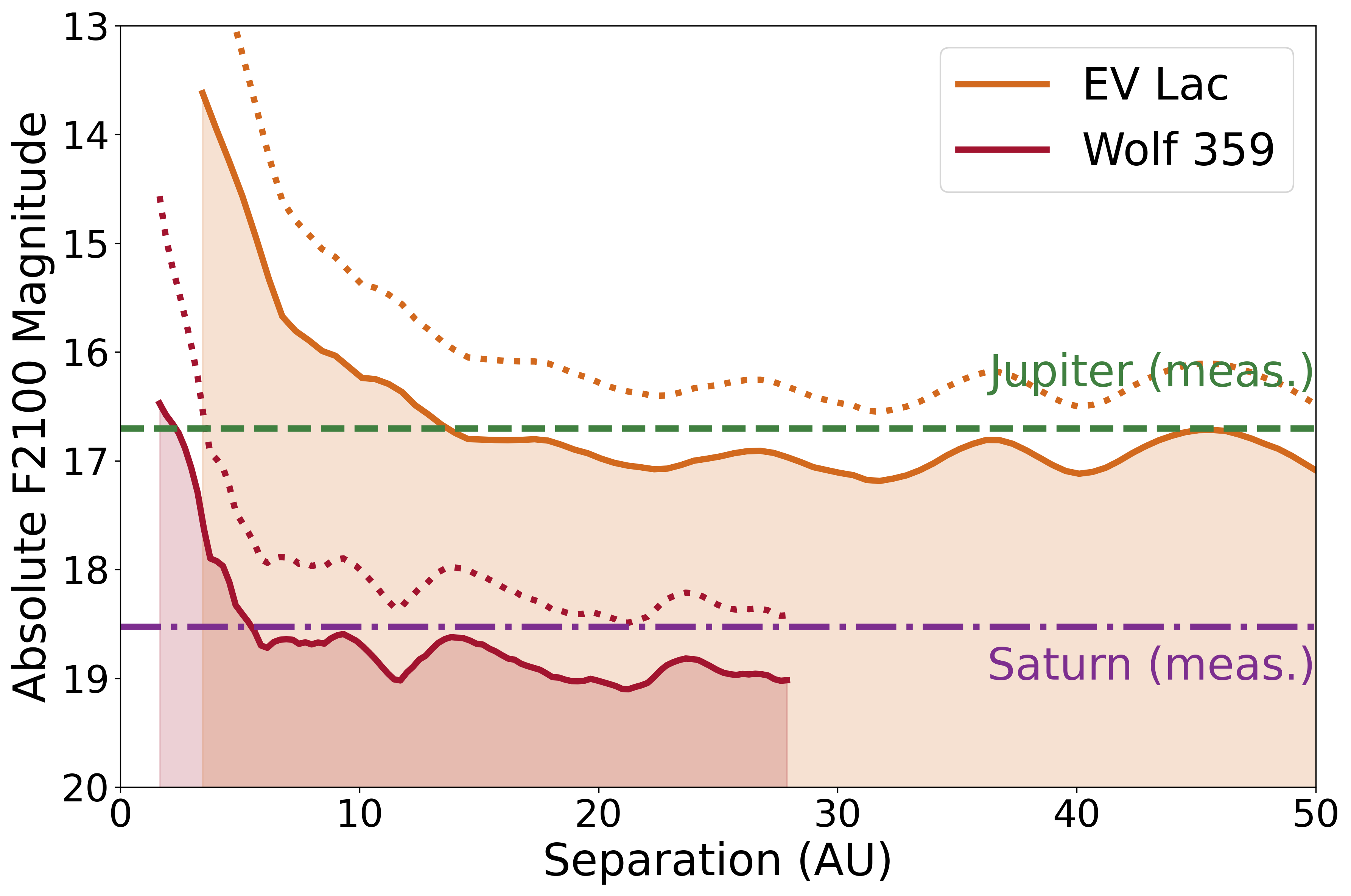}{0.4\textwidth}{(f) MIRI F2100W Imaging Sensitivity in Absolute Magnitude versus AU}}
\caption{\textbf{NIRCam F444W+MASK335R and MIRI F2100W Contrast Curves.} 
The measured 3 S/N (\textit{solid}) and 5 S/N (\textit{dotted}) sensitivity for NIRCam F444W coronagraphy MIRI F2100W imaging are shown for Wolf 359 (\textit{dark blue/red}) and EV Lac (\textit{light blue/orange}).  In approximately 1\,hr of integration, the NIRCam observations reached the 3 S/N background-limited regime at a contrast of $6\times10^{-7}$ by a separation of 2.6 arcsec. The MIRI imaging observations reached the 3 S/N background-limited regime by a similar separation. 
Panels (e) and (f) show the NIRCam and MIRI sensitivity in units of absolute magnitude versus AU as compared to the absolute magnitude of Jupiter (\textit{green dotted line}) and Saturn \textit{(purple dash-dot line)} measured from their spectra. An exoplanet the same magnitude as the measured absolute magnitude of Saturn could be detected orbiting Wolf 359 using MIRI imaging but not NIRCam coronagraphy, highlighting the advantages of using MIRI imaging in this empirically bounded case. }
\label{fig:cc}
\end{figure*}

The contrast curves for the MIRI F2100W observations are shown in the middle row of Figure~\ref{fig:cc}. The contrast curves produced with \texttt{VIP} take into account the necessary small-number statistics corrections at tight inner working angles (see \citealt{Mawet2014}) by applying a student-t correction. Negative fake companion injection was performed to remove flux contributions from potential companions or background sources, following the steps in the \texttt{VIP} tutorial.\footnote{\texttt{VIP} tutorial on negative fake companion injection: \url{https://vip.readthedocs.io/en/latest/tutorials/05A_fm_planets.html}} One source was removed in the Wolf 359 image and two sources were removed in the EV Lac image. The background-limited regime at 3 S/N is reached at $\sim$2.5\arcsec, corresponding to an apparent magnitude limit between $m_{F2100W} = 15.5-16$. Once outside the PSF residuals, the detection is photon-noise limited by the high background flux in F2100W. 

A detector artifact is visible in the MIRI image of Wolf 359 approximately 75 pixels above the central star.  This artifact is visible in the raw images of Wolf 359 and was not due to the PSF subtraction step.  To avoid this issue when estimating the MIRI sensitivity, the contrast curves were created avoiding this section of the image (\texttt{wedge = [180,360]}).

A residual is present in the smallest inner-working angles ($<\sim2 \lambda/D$; 1.2 arcsec) after PSF subtraction which may be due to MIRI's brighter/fatter effect \citep{Argyriou2023}; see center of the MIRI PSF subtracted image in Figure \ref{fig:MIRIraw}.  
This effect introduces subtle variations in the PSF shape for stars with different detector full-well brightnesses, making it challenging to perform precise subtraction for a self-referencing survey where no dedicated reference observations were taken. The brighter/fatter effect takes hold before pixel saturation is reached. 
No mitigation steps were applied in this analysis; however, future correction of the brighter/fatter effect holds promise for enhancing the sensitivity at separations of $<1.2''$. Section~\ref{sec:discussion} discusses potential strategies for mitigating this limitation and improving the performance at small inner-working angles.

Cataloged and uncataloged sources are visible above 3 S/N in both systems. Future proper motion follow-up observations will be required for the confirmation of these sources as background objects or comoving companions.

\section{Results} \label{sec:results}

\subsection{Custom Cold Planet Models \label{sec:planetmodels}}
The atmospheric models currently available for exoplanets and substellar worlds \citep[e.g.][]{Saumon2008, Allard2011, Linder2019, Phillips2020, Marley2021, Lacy2023, Mukherjee2024, Morley2024} do not explore the temperature parameter space required to assess the sensitivity of this program to cold giant planets. We therefore generated custom atmospheric models for objects with effective temperatures as low as 50 K. We used \texttt{PICASO} \citep{Batalha2019, Mukherjee2023}, an open source Python-based atmospheric model to compute one-dimensional pressure--temperature (P-T) profiles in radiative–convective and chemical equilibrium. \texttt{PICASO} has been used to model exoplanet and brown dwarf atmospheres in numerous JWST programs \citep[e.g.][]{Beiler2023, Miles2023, Beiler2024, Biller2024, Lew2024, Petrus2024, Matthews2024} and is rooted in the legacy of the substellar \texttt{EGP} code \citep{Marley1996, MarleyMcKay1999, Fortney2005, Fortney2007, Fortney2008, Robinson2014, Morley2018, Marley2021, Karalidi2021, Morley2024}. 

To incorporate clouds, we followed the \texttt{EddySed} framework described in \citet{Ackerman2001}. \texttt{EddySed} has been widely utilized in various studies \citep[e.g.][]{marley2010, Skemer2016,Morley2015, Rajan2017}, and in suites of models such as those developed by \citet{Saumon2008} and the Sonora Diamondback models \citep{Morley2024}. We have implemented new developments within \texttt{PICASO} to self-consistently treat clouds using \texttt{Virga} \citep{virga}, a Python-based implementation of the \citet{Ackerman2001} model. Water-ice clouds form when water vapor exceeds the saturation vapor pressure above the atmospheric layer where the P–T profile intersects the condensation curve. These clouds are parameterized in \texttt{Virga} by the sedimentation efficiency parameter, $f_{\rm sed}$, which governs their vertical extent. Higher $f_{\rm sed}$ values produce optically thinner, vertically constrained clouds with large particles, whereas lower $f_{\rm sed}$ values result in optically thicker, more vertically extended clouds with smaller particles.

For this survey, we generated a small grid of models with both clear and fully cloudy atmospheres. This grid spanned effective temperatures of $T_{\rm eff}$ = [50 - 300 K] at 50 K increments, surface gravities of log(g) = [3.0 - 4.75] at 0.25 dex (cgs), a metallicity of [M/H] = +0.5 (relative to the sun),  and a solar C/O ratio (0.458, \cite{Lodders2010}). The cloudy models used $f_{\rm sed} = 8$ across all temperatures, while additional models with $f_{\rm sed} = 4$ and $6$ only go up to models with $T_{\rm eff} \leq 125$ K due to challenges at warmer temperatures with convergence. In all cases, H$_2$O is the only condensing species. Future work will explore additional condensates, such as NH$_3$ and CH$_4$, which are expected to form in the coldest atmospheres. These custom models are a subset of a larger grid that will be published in the future to extend Sonora Bobcat \citep{Marley2021}, Sonora Diamondback \citep{Morley2024}, and Sonora Elf Owl \citep{Mukherjee2024} to lower temperatures and a broader range of metallicities and C/O ratios along with updated evolutionary tracks.

Using these 1D thermal structures, we computed moderate-resolution thermal emission spectra with \texttt{PICASO}. These spectra spanned 0.3 to 28 $\upmu$m at a resolving power of 5000. Opacities are taken from \citet{Freedman2008} with updates from \citet{Freedman2014}, and we incorporate the revised opacity tables from \citet{Morley2024} and \citet{Mukherjee2024} which include improvements to H$_2$O \citep{Polyansky2018}, CH$_4$ \citep{Hargreaves2022}, FeH \citep{Hargreaves2010}, and alkali metal opacities.

\subsection{Anchors from Solar System planets}

\begin{figure*}
\gridline{\fig{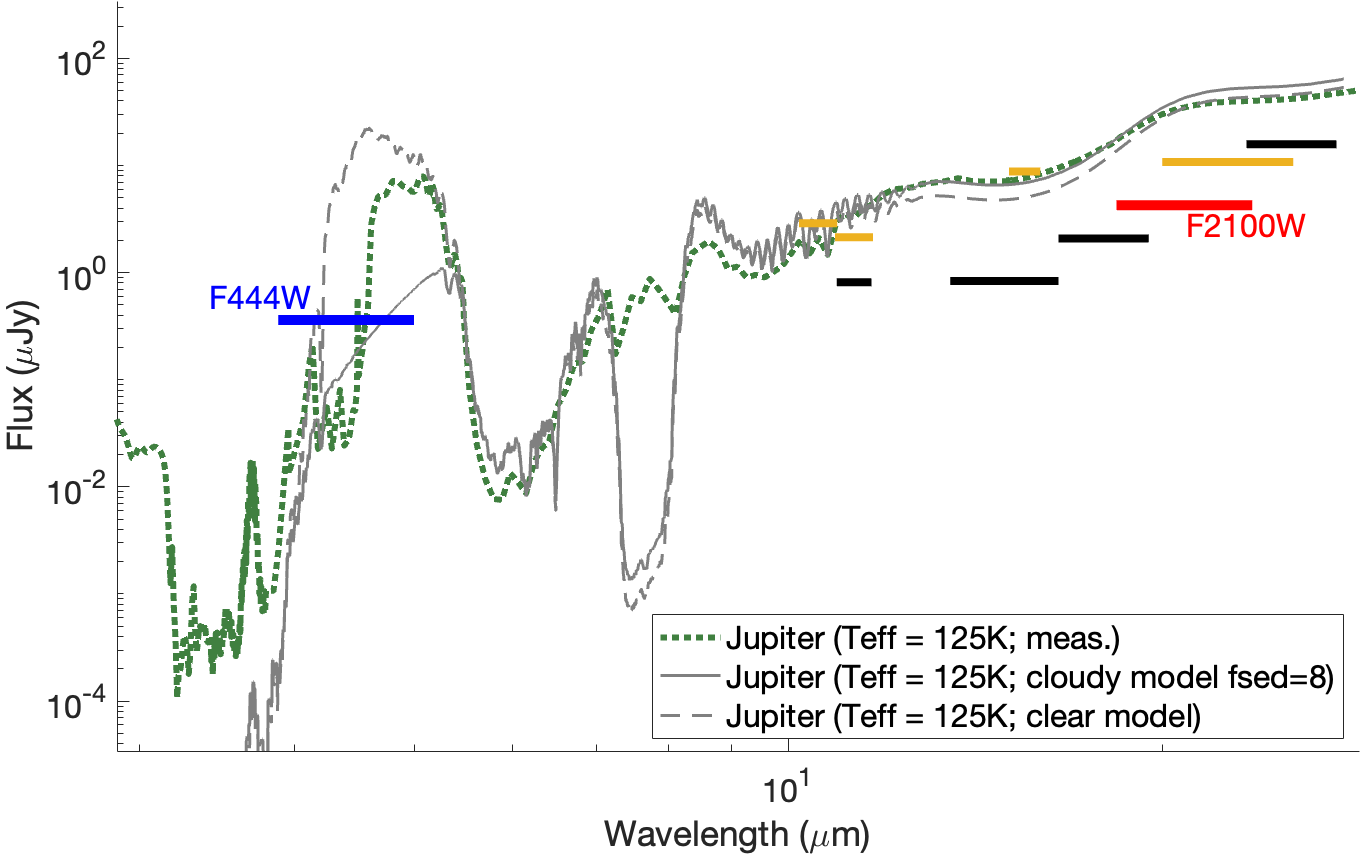}{0.5\textwidth}{(a) Flux of Jupiter}
\fig{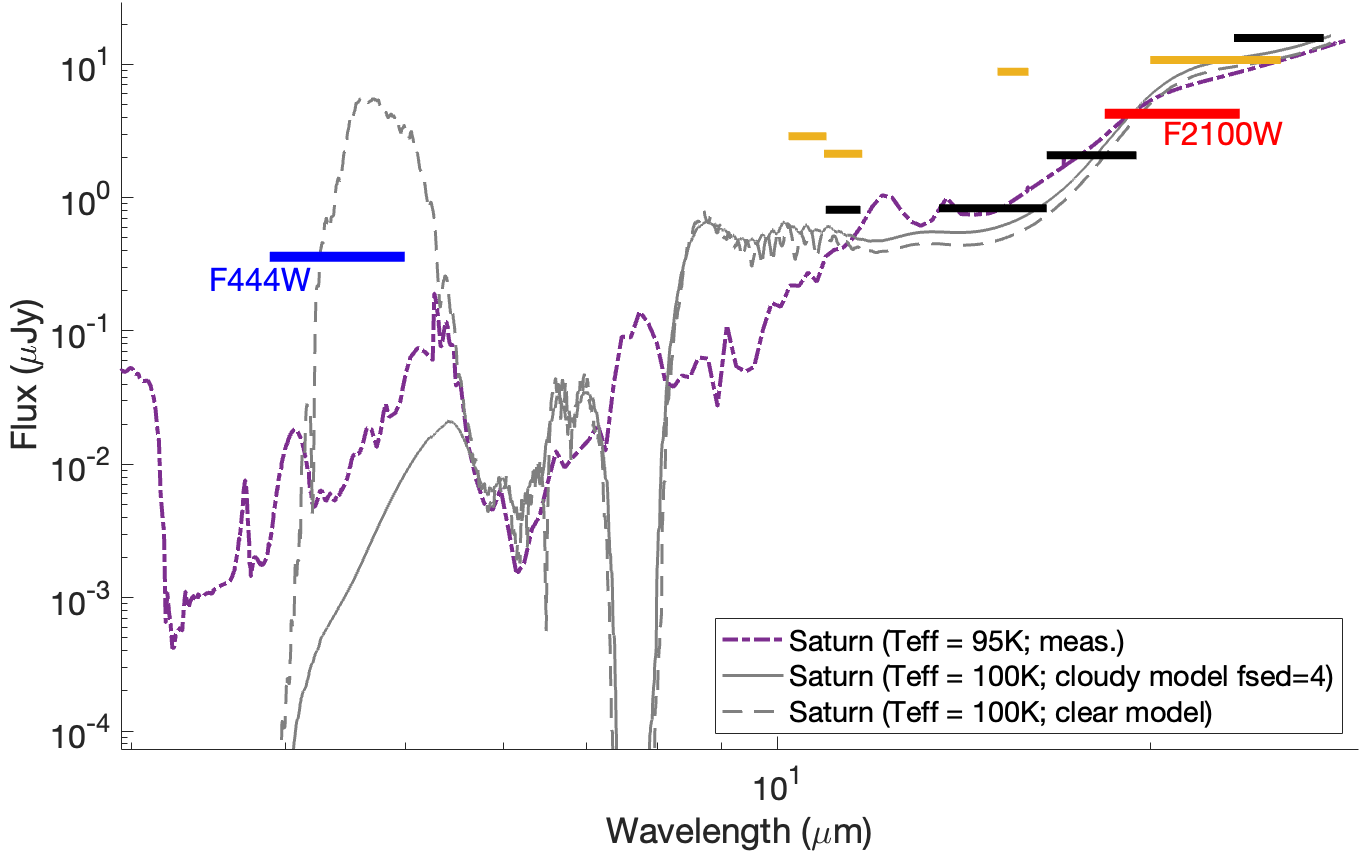}{0.5\textwidth}{(b) Flux of Saturn }}
\caption{\textbf{Comparison of Jupiter/Saturn measured spectra with clear and cloudy equilibrium models.} The measured spectra of Jupiter and Saturn were scaled to the distance of Wolf 359 (2.4 pc) for comparison with clear and cloudy models. The horizontal lines represent detection limits for various JWST instrument modes: NIRCam F444W coronagraphy (\textit{blue}), MIRI F2100W imaging (\textit{red}), MIRI coronagraphy (\textit{gold}), and MIRI imaging (\textit{black}). When the spectra exceed the instrument limit, the flux can be detected using that mode. Most JWST modes could detect a true Jovian-analog orbiting Wolf 359 at separations in the background-limited region of the contrast curve. However, when considering the case of a true Saturn-analog, NIRCam F444W coronagraphy is only capable of detecting 
a Saturn-temperature exoplanet if the planet has a clear atmosphere.} 
\label{fig:spectracompare}
\end{figure*}

\begin{figure*}
    \centering
\includegraphics[width=0.85\linewidth]{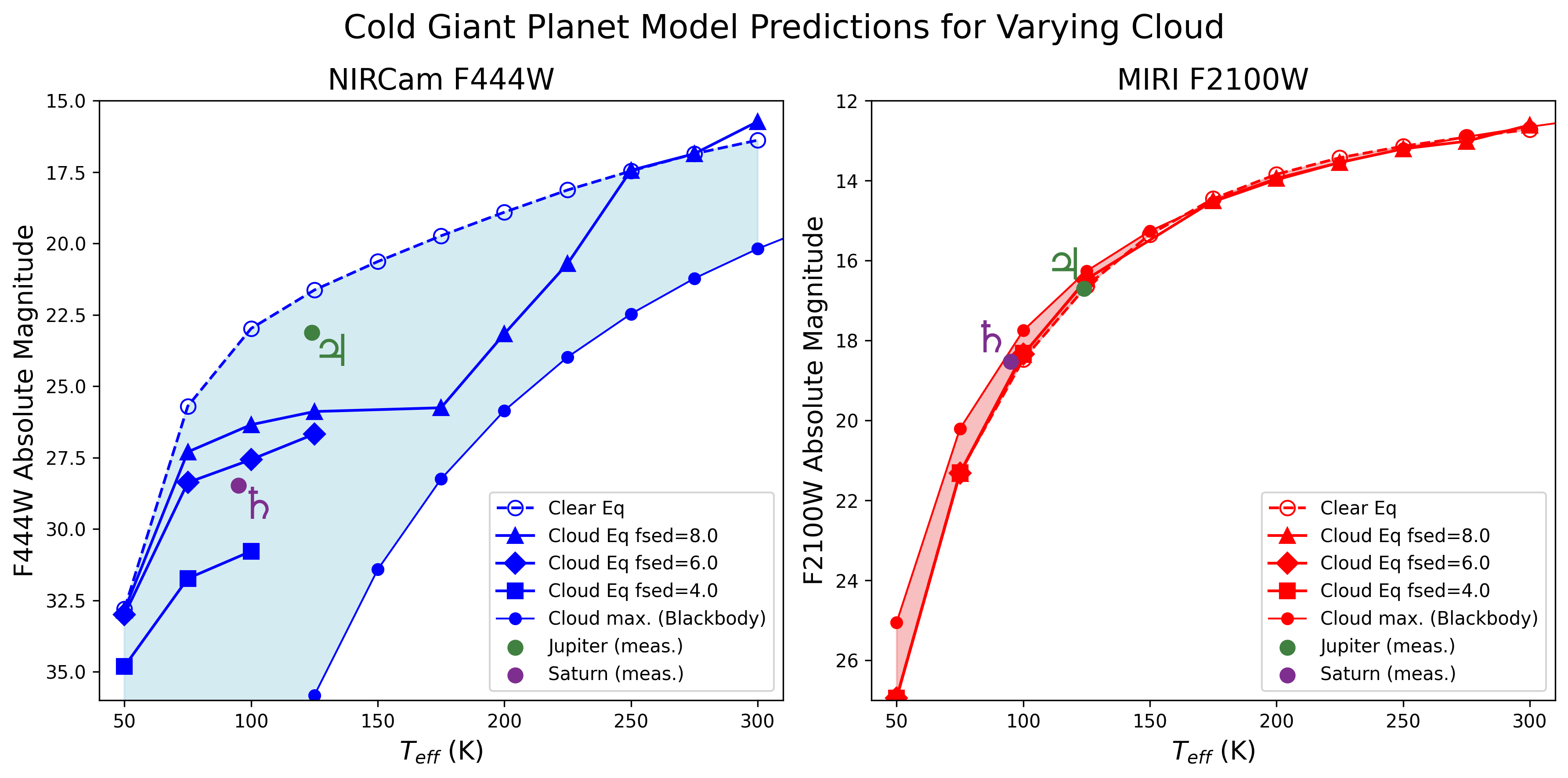}
    \caption{\textbf{Planet brightness estimates for  clear and cloudy cold giant exoplanets using equilibrium models:} The absolute magnitude estimates for clear atmosphere (\textit{dashed line, open circle}) and cloudy atmosphere (\textit{solid line, filled shapes}) are shown for NIRCam F444W coronagraphy (\textit{left in blue}) and MIRI F2100W imaging (\textit{right in red}).  The absolute magnitudes were estimated using the custom cold giant planet models described in Section \ref{sec:planetmodels}. The maximum-cloud model magnitudes (\textit{solid line circle}) were computed using the SED of a blackbody.  While the magnitude at F2100W is equally capable of detecting clear and cloudy planets, the F444W magnitude varies significantly depending on cloud conditions.} 
    \label{fig:comparemagsofmodels}
\end{figure*}

To ground the cold planet models using real atmospheric data, we compared our models to the observed spectral energy distributions (SEDs) of Jupiter and Saturn.  
We compiled data for the Jupiter and Saturn spectra from two sources: (1) the JWST Exposure Time Calculator (ETC) spectra of Jupiter and Saturn\footnote{\url{https://www.stsci.edu/hst/instrumentation/reference-data-for-calibration-and-tools/astronomical-catalogs/solar-system-objects-spectra}}, and (2) the spectra from \cite{2016PASP..128a8005N}. A composite spectrum was necessary, as the \cite{2016PASP..128a8005N} data do not extend to the longest wavelengths, while the ETC documentation notes that its spectra do not reliably represent the 4-8\,$\mu$m region. To address this, we adopted the ETC spectrum except in the 4-8\,$\mu$m range, where we substitute the data from \cite{2016PASP..128a8005N}.
Because the separation of Jupiter and Saturn to Earth were not stated in the literature where the spectra data originated, we anchored the flux of the Jupiter and Saturn spectra to our models at 20\,$\mu$m where the planet flux is agnostic to atmospheric conditions in this temperature regime (see Figure \ref{fig:comparemagsofmodels} as well as Eps Indi Ab models in \citealt{Matthews2024}). 
We show a comparison of the measured Jupiter and Saturn spectra with a clear and cloudy models at the corresponding effective temperature in Figure \ref{fig:spectracompare}.  The flux was scaled as if the planet/model is at the distance equivalent to that of Wolf 359 (2.41\,pc) in Figure \ref{fig:spectracompare}.

 The MIRI imaging sensitivity limits plotted for in Figure \ref{fig:spectracompare} correspond to the background values provided in the JWST STScI documentation.\footnote{JWST STScI documentation: \url{https://jwst-docs.stsci.edu/jwst-mid-infrared-instrument/miri-performance/miri-sensitivity\#gsc.tab=0}} 
At F2100W, the documented sensitivity closely matches the background-limited performance achieved by GO 6122, suggesting that these values are representative for performing high-contrast imaging with the other MIRI imaging filters. The coronagraphic sensitivity limits plotted in Figure \ref{fig:spectracompare} are approximate, based on estimates from the JWST Helpdesk, and are generally less sensitive than the imaging limits due to reduced throughput and narrower bandwidths. For this reason, MIRI imaging may be optimal over MIRI coronagraphy but this trade space requires further investigation.

We find that the measured spectra of Jupiter and Saturn are best reproduced by a combination of clear and cloudy atmospheric models. Jupiter’s flux at F444W (2.20\,$\mu$Jy) corresponding to $T_{\mathrm{eff}} = 124.4 \pm 0.3$\,K \citep{Roman2023} is well matched by a blend of 80.8\% of the 125\,K-cloudy model  with $f_{\mathrm{sed}} = 8$ and 19.2\% of the 125\,K-clear model. Similarly, Saturn’s measured F444W flux (0.0146\,$\mu$Jy) at $T_{\mathrm{eff}} = 95.0 \pm 0.4$K \citep{Roman2023} is best described by 99.5\% of the 100\,K-cloudy model with $f_{\mathrm{sed}} = 4$ and 0.5\% of the 100\,K-clear model.
Although no exoplanets currently have reported $f_{\mathrm{sed}}$ values in this low-temperature regime, the directly imaged companion Eps Ind Ab shows evidence for flux suppression that may be caused by cloudiness \citep{Matthews2024}. The newly reported exoplanet candidate TWA 7\,b ($\sim$340\,K) has a measured $f_{\mathrm{sed}}$ of $3.1^{+0.6}_{-0.4}$ \citep{Lagrange2025}. These comparisons suggest that clear atmospheric models alone are insufficient to accurately describe the emergent flux of cold giant planets.

\subsection{Survey Sensitivity to Imaging Cold Planets around Wolf 359 and EV Lac}

To compare the NIRCam and MIRI sensitivity in terms of the detectable planetary effective temperature ($T_{eff}$), we applied the atmospheric models described in Section \ref{sec:planetmodels} to convert the contrast curves in Figure \ref{fig:cc} into limits on effective temperture. We computed the expected magnitudes in F444W and F2100W using the Pandeia Engine \citep{Pandeia} assuming a planetary radius of $1\,R_{\mathrm{Jup}}$. When models with multiple surface gravity values were available, we calculated the flux by averaging the outputs of all models at a given temperature.

Figure \ref{fig:comparemagsofmodels} shows the estimated F444W and F2100W magnitudes for a cold exoplanet at each effective temperature. We included cloudy model solutions with $f_{\text{sed}}$ values of 4, 6, and 8, along with two bounding cases: clear and "maximally cloudy." As $f_{\text{sed}}$ decreases, indicating increased cloudiness, the models converge towards blackbodies. The "maximally cloudy" case was thus approximated by a blackbody at the respective temperature. While real planets are not blackbodies, this approximation effectively bounds their expected flux range, as actual planets should generally have flux values between those of a blackbody and a clear model. These models show little variation in flux in the F2100W filter ($ \lesssim 0.4\times$ variation in flux). However, at F444W, we find that clouds can significantly reduce the flux. In extreme cases, the magnitude difference can exceed 10 magnitudes ($>10,000\times$ in flux). As a result, F2100W provides a more precise anchor to measure of a planet's temperature and mass than F444W.

\begin{figure*}
    \centering
    \includegraphics[width=0.95\linewidth]{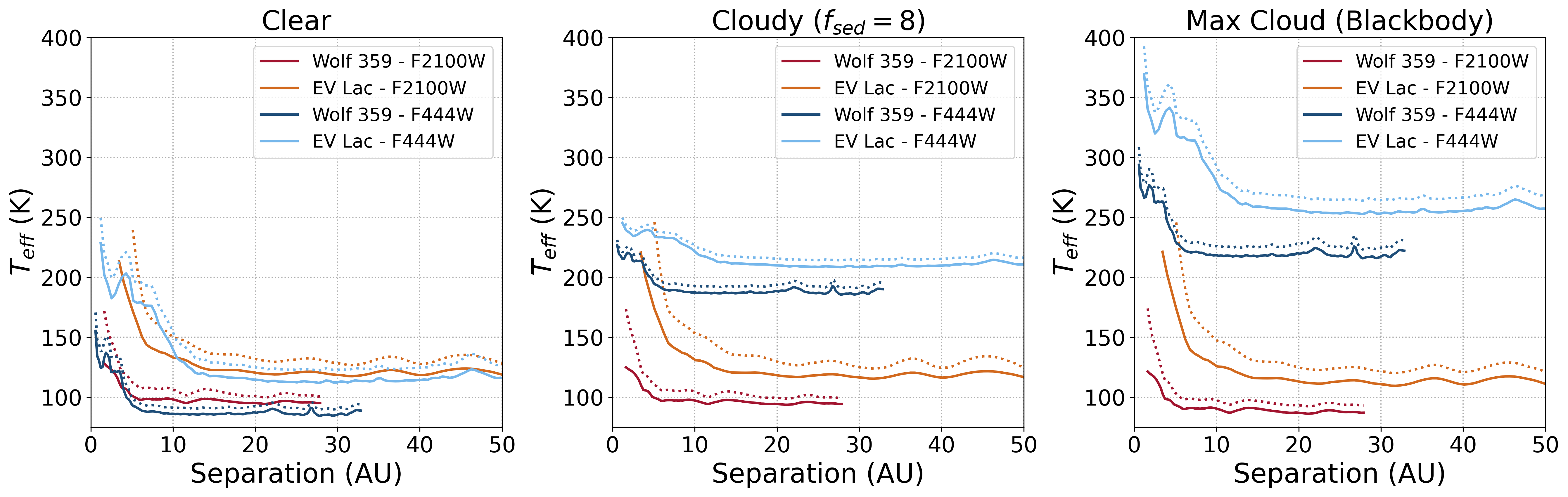}
    \caption{\textbf{GO 6122 Sensitivity to Cold Planets Assuming 3 Varying Atmospheric Conditions}. We converted our measured 3 S/N \textit{(solid)} and 5 S/N \textit{(dotted)} contrast curves into estimates of the effective temperatures of a potentially detectable exoplanet. We adopt three atmosphere cases: clear \textit{(left)}, cloudy equilibrium with $f_{sed} = 8$ \textit{(middle)}, and maximum cloud  \textit{(right)}.  While the limits for MIRI F2100W imaging and NIRCam F444W coronagraphy are similar assuming a planet with a clear atmosphere, MIRI F2100W imaging can be more sensitive to cold planets by more than 100K if the planet has a cloudy atmosphere. We caution that the $f_{sed} = 8$ cloudy equilibrium model set is only mildly cloudy — both Saturn and TWA 7b are significantly cloudier ($f_{sed} = 3 - 4$; \citealt{Lagrange2025}). 
    As a result, the middle plot may be overly optimistic for NIRCam and not fully representative of typical cloudiness levels on cold planets.}
    \label{fig:CCbyTeff}
\end{figure*}

Figure \ref{fig:CCbyTeff} shows our measured contrast curves for Wolf 359 and EV Lac translated to the coldest-detectable planet using three atmospheric models: clear, cloudy ($f_{\text{sed}} = 8$), and maximally cloudy (blackbody).  MIRI 21$\mu$m imaging can detect companions as cold as 94\,K for Wolf 359 (colder than Saturn, $T_{\rm eff}$ = 95\,K) and 114 K for EV Lac (colder than Jupiter, $T_{\rm eff}$ = 124\,K) in all atmospheric cases. For Wolf 359, sub-100 K planets can be detected beyond 4.8 AU suggesting that a planet with the same temperature (95 K) and orbital separation (9.5 AU) as Saturn could be detected. Given that Wolf 359 is less than one-third the age of our Solar System, this temperature limit corresponds to sub-Saturn masses including masses in the ice-giant regime. 

In the clear-atmosphere case, the limits for MIRI imaging and NIRCam coronagraphy are within 10\,K for the two stars. For example, the coldest temperature planet detectable to 3 S/N in the clear atmosphere case for the EV Lac system is predicted to be 114 K using F2100W imaging and 111 K using F444W coronagraphy.  
However, for the $f_{sed} = 8$ cloudy models, MIRI F2100W imaging can detect exoplanets that are up to 90\,K colder than those accessible with NIRCam
(limit of $T_{\rm eff}\,F2100W = 93$ K versus $T_{\rm eff}\,F444W = 185$ K for Wolf 359; gap is larger for EV Lac). In the case of maximum clouds, the performance gap between the two  modes increases to be greater than 130 K (limit of $T_{\rm eff}\,F2100W = 84$\,K versus $T_{\rm eff}\,F444W = 215$\,K for Wolf 359; gap larger for EV Lac).

Overall, we find that MIRI F2100W imaging outperforms NIRCam F444W coronagraphy in detecting exoplanets <300 K in these two systems ($<6$\,pc). This advantage holds at both small inner working angles and in the background-limited regime. For example, in the clear-atmosphere case (which is optimal for F444W) shown in panel 1 of Figure \ref{fig:CCbyTeff}, F2100W imaging and F444W coronagraphy performance was comparable across most separations. Additionally, MIRI F2100W offers a more precise photometric band for measuring planet temperatures as it is less affected by atmospheric conditions. When used in conjunction, NIRCam F444W coronagraphy and MIRI F2100W imaging can provide an initial probe of the atmosphere type of these cold worlds.

\section{Discussion} \label{sec:discussion}

\subsection{Future opportunity for improving the sensitivity}

While the analysis presented in this work is sufficient to show that JWST is capable of directly studying Jupiter and Saturn analog exoplanets, future JWST observations and on-going efforts to enhance the data reduction techniques could open the door to improved sensitivities beyond those presented. For this analysis, we elected to limit the number of reference stars used in the RDI PSF subtraction to only include images collected as part of the initial observations of the GO 6122 program. The remaining observations for GO 6122 along with the future Cycle 4 program SURVEY 8581: \textit{Hot On The Hunt for frigid exoplanets}\footnote{JWST SURVEY 8581: \url{https://www.stsci.edu/jwst-program-info/program/?program=8581} } will obtain tens of additional PSFs for MIRI imaging at F2100W. 
Future analyses could improve the sensitivity limits by including additional reference stars as images become publicly available. 
The high-contrast imaging community is currently taking steps to build a PSF library framework for \texttt{spaceKLIP} NIRCam coronagraphic analysis.  These efforts are motivated by past successes in improving high-contrast imaging sensitivities with ground-based instruments \citep{Sanghi2024,Xie2022,Xuan2018} and the Hubble Space Telescope \citep{Choquet2016,Sanghi2022}.

The MIRI observing community is also actively exploring ways to mitigate the non-linearity effects caused by MIRI's brighter-fatter effect \citep{Argyriou2023}. This effect occurs when variations in accumulated charge between neighboring pixels cause charge to diffuse into pixels with lower accumulated charge. As a result, the PSF shape of stars subtly changes with brightness, introducing residuals near the core during the PSF subtraction step. This issue affected the sensitivity at the smallest inner-working angles in this analysis ($<1.2$\,arcsec).

One strategy to negate the brighter-fatter effect is to more closely pair future science observation with reference stars that better balance the detector counts. This can be accomplished by having a dedicated reference star as part of an observation, but this strategy comes at the cost of increasing the overhead time per science target. 
If including a reference star in the observation is not possible, \cite{Valentine2024} and \cite{Grant2023} demonstrated a promising strategy  to mitigate the brighter-fatter effect in post-processing for MIRI-LRS. In their method, they calibrated the mis-matched reference image by modifying the ramp-fitting step in Stage 1 of the JWST pipeline. They re-fit their reference frame to artificially create the shape of the dimmer science image using a select set of groups (i.e. groups 12 - 40 were used for fitting out of the 175 groups in the reference star image).

If the brighter-fatter effect is mitigated, MIRI F2100W imaging should achieve a significantly improved inner working angle. \cite{Limbach2024} found that in cases where brighter-fatter effects are not a concern, direct detection of exoplanets as close as 0.3$\lambda/D$ (0.2'') is possible at F2100W. This is smaller than the typically quoted inner working angle of the NIRCam coronagraphy masks, although the throughput of these masks is gradual and these inner working angles have been achieved with the wedge mask \citep{2025AJ....169..209B}. This suggests that if brighter-fatter effects can be mitigated, MIRI imaging may offer an advantage at small inner working angle compared to NIRCam coronagraphy while achieving similar or favorable planet sensitivity.

\subsection{Predicted sensitivity to imaging cold exoplanets in the 70pc sample}

\begin{figure*}
    \centering
    \includegraphics[width=1\linewidth]{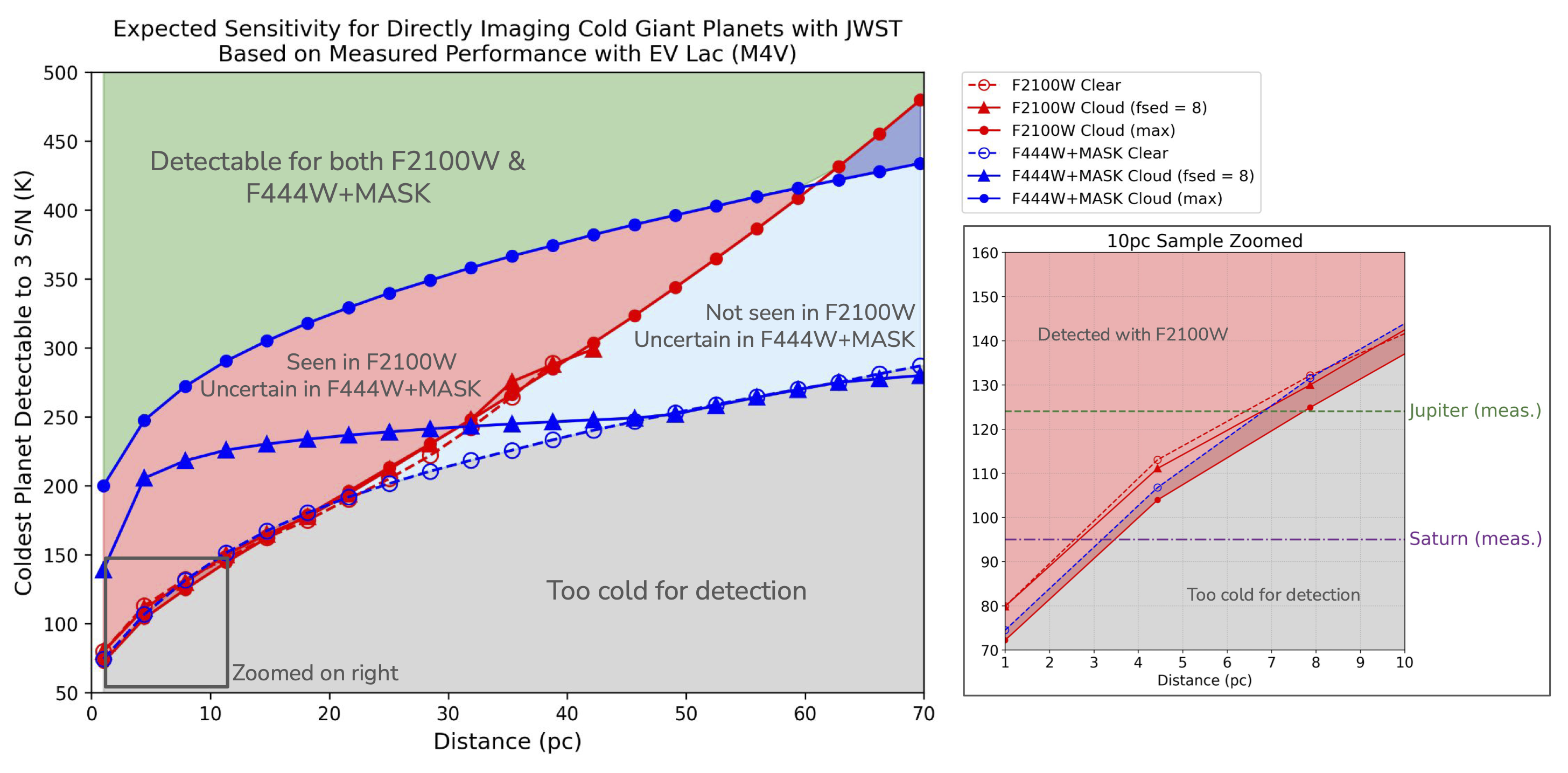}
 \caption{\textbf{Coldest Planet Detectable by Distance to 3 S/N.} 
To evaluate the tradespace between F444W NIRCam coronagraphy and F2100W MIRI imaging, we compared the coldest planet detectable in the background-limited regime at 3 S/N for both modes out to 70\,pc.
This plot was constructed using the measured background limits in apparent magnitude for EV Lac (M4V): F444W = 20.8 and F2100W = 15.75. The blue lines represent the detection limits for NIRCam F444W+MASK335, while the red lines indicate the limits for MIRI F2100W. These limits are shown for three atmospheric cases in equilibrium: clear, cloudy ($f_{sed} = 8$), and maximally cloudy.
Planets in the gray region fall below all detection limits and cannot be observed with either instrument mode. Planets in the green region are hot enough to be detected by both modes, regardless of their atmospheric properties. The red region marks planets that are always detectable with MIRI F2100W but only detectable with NIRCam F444W+MASK if they have a clear atmosphere. The light blue region contains planets undetectable with MIRI F2100W but potentially detectable with NIRCam F444W+MASK if they have a clear atmosphere. The dark blue region marks planets too faint for MIRI F2100W but bright enough to always be detected with NIRCam F444W+MASK, regardless of their atmospheric properties.
MIRI F2100W provides a detection advantage that is independent of atmospheric conditions for systems within 20\,pc. A Jupiter-temperature planet can be detected using MIRI F2100W within 7\,pc while a Saturn-temperature planet can be detected within 3\,pc.} 
    \label{fig:Teffbydist}
\end{figure*}

Although we find that MIRI F2100W imaging is advantageous for directly imaging cold exoplanets in nearby systems like those in the GO 6122 program ($<6$\,pc), the trade space will change between these observing modes for systems with less proximity. 
We use Figure \ref{fig:Teffbydist} to explore the trade space between NIRCam F444W coronagraphy and MIRI F2100W imaging for systems out to 70\,pc. This figure plots the coldest planet detectable to 3 S/N based on our measured apparent magnitude limits reached for EV Lac (apparent magnitude in background regime: F444W = 20.8, F2100W = 15.75).   
Figure \ref{fig:Teffbydist} is representative of stars with a similar spectral type to EV Lac (M4V).

At greater distances, the coldest planets will not emit enough flux to exceed the image 
background noise, as the apparent magnitude decreases with distance while the background noise remains constant. F444W observations become more advantageous for detecting hotter planets as a larger fraction of the planet flux falls within the 4–5 micron range. According to Wien's displacement law, the peak wavelength of blackbody radiation is 4.5 microns for a 644\,K object and 21 microns for a 138\,K object.

We find that the parameter space can be divided into five regions in Figure \ref{fig:Teffbydist} based on which observing mode is advantageous at what proximity. The region where a planet would be too cold for detection by either of the GO 6122 observing-mode configurations is shaded in gray. 
We find that no planets can be detected at any distance beyond Proxima Centauri below 70 K with the GO 6122 observing modes, approaching the effective temperature of Uranus and Neptune (T$_{eff} \approx 60$\,K; \citealt{Roman2023}). Determining whether longer wavelength MIRI imaging (F2550W) or other observing modes could provide an avenue towards imaging planets colder than 70\,K in the very nearest systems may be a worthwhile exploration but beyond the scope of this work.  

The region shaded in red represents the parameter space where a companion would be detected to 3 S/N with MIRI F2100W imaging but not necessarily NIRCam F444 coronagraphy depending on the atmospheric conditions on the planet.  At a distance of approximately 22\,pc and a planet temperature of 190 K, a transition emerges where a planet with a clear atmosphere could be detectable with NIRCam F444W coronagraphy but will not be detectable with the MIRI F2100W imaging configuration (region shaded in light blue). While companions in the green shaded region could be detectable with either observing mode, those cold enough to lie in the region shaded dark blue will only be detectable through NIRCam F444W coronagraphy. This transition occurs at approximately 61\,pc above planet temperatures of 420\,K.
Overall, we find that when the goal of a JWST direct imaging program is to reveal as many giant exoplanets as possible colder than 500\,K, targets with M-dwarf host stars within 22\,pc are best observed with MIRI F2100W imaging. Systems beyond 61\,pc are best observed with NIRCam F444W coronagraphy. 

Examining the trade space for the 10\,pc sample, we find that a Jupiter-temperature planet (124\,K; \citealt{Roman2023}) is detectable using MIRI F2100W imaging for systems within 7\,pc. Using NIRCam F444W coronagraphy would not guarantee such a detection of companions colder than Jupiter unless the exoplanet had a clear atmosphere.  Similarly, a Saturn-temperature exoplanet (95\,K; \citealt{Roman2023}) is detectable within approximately 3\,pc using MIRI F2100W imaging but not F444W coronagraphy. All stars in the GO 6122 program have the proximity needed to find exoplanets colder than Jupiter. Two stars in the survey (Wolf 359 and Ross 154) have proximity close enough to reveal companions colder than Saturn.

\section{Conclusion} \label{sec:conclusion}

In this work, we demonstrated that JWST can directly image exoplanets with temperatures, ages, and orbital separations comparable to the gas giants in our own Solar System. Using initial observations from the GO 6122 program with NIRCam F444W coronagraphy and MIRI F2100W imaging, we analyzed the detectability of cold exoplanets orbiting Wolf 359 and EV Lac. We interpreted the measured sensitivity using custom atmospheric models spanning $50$–$500$\,K and compared it against the empirically measured spectra of Jupiter and Saturn. We  extended this comparison to estimate the performance of NIRCam coronagraphy and MIRI imaging for systems out to 70\,pc.

Our findings and recommendations for future JWST programs targeting cold ($\lesssim350$\,K) exoplanets and Solar System analogs are summarized as follows:

\begin{itemize}

\item  \textbf{JWST has the sensitivity to detect Jupiter/Saturn analogs in nearby systems}: The GO 6122 observations show that direct imaging with JWST is capable of revealing cold giant exoplanets with temperatures that are analogous to our own solar system giant planets ($< 125$K) at the orbital separations we expect them to be common (5-15 AU).  For the very nearest systems ($<3$\,pc), JWST is capable of revealing planets at temperatures of less than 100 K in the background limited regime ($>2.5$\,arcsec).

    \item \textbf{Selecting planet models that incorporate clouds/flux suppression at F444W is necessary to accurately describe cold giant planets:} 
    Cold exoplanets ($\lesssim350$\,K) are expected to have cloud cover that suppresses flux in the 3.5–5\,$\mu$m region. Clear-atmosphere models systematically overestimate detectability at F444W. Incorporating clouds into models — along with other potential flux-suppressing effects like disequilibrium chemistry or high metallicity — is essential for realistic yield predictions and observing program design.

    \item \textbf{
   MIRI F2100W imaging is advantageous over NIRCam F444W when detecting the coldest planets ($\lesssim250$\,K) in nearby systems ($<20$\,pc)}: MIRI F2100W imaging can detect cold planets that NIRCam cannot detect.  Because the F2100W flux band is less sensitive to atmospheric conditions, MIRI F2100W measurements also enable more robust temperature/mass estimates. NIRCam F444W measurements suffer from degeneracies with cloud properties leading to large uncertainties.

     \item \textbf{NIRCam F444W becomes advantageous at larger distances ($\gtrsim60$\,pc)}: At greater distances, detectable planets are generally hotter, and a larger fraction of their flux fall in the 3.5–5\,$\mu$m range. As a result, the relative benefit of MIRI imaging diminishes beyond $\sim$60\,pc.

    \item\textbf{NIRCam and MIRI observations are complementary:} If performed together, these observations provide a pathway not only for the discovery of new worlds but also in their initial characterization. MIRI photometry can anchor the temperature/mass measurement while NIRCam photometry can probe the clouds/disequilibrium chemistry/metallicity of the world.

\end{itemize}

Although NIRCam corongraphy is currently more widely adopted within the JWST direct imaging community, MIRI F2100W imaging shows promise in opening a new window to detecting and characterizing cold, mature exoplanets at Solar System–like separations. We recommend that future JWST surveys seeking to detect cold giant planets strongly consider incorporating MIRI imaging. While this study focused on two observing modes, our results suggest that MIRI F2100W will outperform other modes in key regions of distance, separation, and temperature space.

The strategies outlined in this work point the way towards uncovering a potentially substantial population of cold, low-mass giant exoplanets and future direct imaging of these planets offers a path to placing our own solar system in the broader context of planetary systems throughout the galaxy.

\section{ACKNOWLEDGEMENTS}

We would like to acknowledge the thousands of people who dedicated themselves to the design, construction, and commissioning of JWST.  The authors would also like to thank Phil Hinz, Sagnick Mukherjee, and Steve Ertel for their early contributions to the development of the GO 6122 program as well as Mark Marley and Jonathan Fortney for their support roles in the modeling efforts. 
The authors would also like to acknowledge the following staff at Space Telescope who contributed to the planning and execution of the observations: Julian Girard, Marshall Perrin, Weston Eck, Jonathan Aguilar, and David Golimowski.

This work is based on observations made with the NASA/ESA/CSA James Webb Space Telescope. The data were obtained from the Mikulski Archive for Space Telescopes at the Space Telescope Science Institute, which is operated by the Association of Universities for Research in Astronomy, Inc., under NASA contract NAS 5-03127 for JWST. These observations are associated with JWST program GO 6122 whose support was provided by NASA through a grant from the Space Telescope Science Institute which is operated by the Association of Universities for Research in Astronomy, Inc., under NASA contract NAS 5-03127. C.V.M and J.M. acknowledge the additional grant support of STScI grants JWST-AR-01977.004-A, JWST-GO-02327.010-A, JWST-AR-03245.004-A, JWST-GO-03337.004-A, and JWST-GO-04050.009-A. C.V.M acknowledges support of NASA XRP grant 80NSSC24K0958.

\software{VIP: Vortex Imaging Processing python package \citep{VIP}, spaceKLIP \citep{Kammerer2022,Carter2023_hip65426}, Species \citep{Stolker2020species}, PanCAKE \citep{Girard2018, Perrin2018, CarterPancake}, PyKLIP \citep{PyKLIP}, Pandeia \citep{Pandeia}, \texttt{PICASO} \citep{Batalha2019, Mukherjee2023}, Astropy \citep{astropy:2013, astropy:2018, astropy:2022}. ChatGPT was used to improve the wording at the sentence level and to assist in coding.}

\bibliography{sample631}{}
\bibliographystyle{aasjournal}

\end{document}